\documentclass[a4paper]{spie}  
\usepackage[]{graphicx}

\title{Fundamental Performance Determining Factors of the Ultrahigh-Precision Space-Borne Optical Metrology System for the LISA Pathfinder mission}

\author{Gerald Hechenblaikner and Reinhold Flatscher
\skiplinehalf
EADS Astrium, 88039 Friedrichshafen, Germany
}

\authorinfo{Further author information: (Send correspondence to G.H.)\\G.H.: E-mail: Gerald.Hechenblaikner [at] astrium.eads.net, Telephone: +49 7545 82435\\  R.F.: E-mail: Reinhold.Flatscher [at] astrium.eads.net, Telephone: +49 7545 84979}

\begin{document}
\maketitle

\section*{}
This paper has been published {\it Modeling Aspects in Optical Metrology IV}, Bernd Bodermann, Karsten Frenner, Richard M. Silver, Editors, Proc. SPIE 8789, 87890X (2013), and is made available as an electronic preprint with permission of SPIE.
Copyright 2013 Society of Photo Optical Instrumentation Engineers. One print or electronic copy may be made for personal use only. Systematic reproduction and distribution, duplication of any material in this paper for a fee or for commercial purposes, or modification of the content of the paper are prohibited.

\begin{abstract}
The LISA Pathfinder mission to space employs an optical metrology system (OMS) at its core to measure the distance and attitude between two freely floating test-masses to picometer and nanorad accuracy, respectively, within the measurement band of [1 mHz, 30 mHz]. The OMS is based upon an ultra-stable optical bench with 4 heterodyne interferometers from which interference signals are read-out and processed by a digital phase-meter. Laser frequency noise, power fluctuations and optical path-length variations are suppressed to uncritical levels by dedicated control loops so that the measurement performance approaches the sensor limit imposed by the phase-meter. The system design is such that low frequency common mode noise which affects the read-out phase of all four interferometers is generally well suppressed by subtraction of a reference phase from the other interferometer signals. However, high frequency noise directly affects measurement performance and its common mode rejection depends strongly on the relative signal phases. We discuss how the data from recent test campaigns point towards high frequency phase noise as a likely performance limiting factor which explains some important performance features.
\end{abstract}

\keywords{optical metrology, heterodyne interferometry, phase noise, quantization noise, space mission}

\section{INTRODUCTION}
\label{sec:intro}  
The LISA Pathfinder Mission\cite{Vit2009} is a technological precursor to the Laser-Interferometer Space Antenna (LISA) mission which aims to detect gravitational waves with interferometry\cite{Bel2008}.
One of the subsystems at its core, the Optical Metrology System (OMS), performs interferometer measurements of position and attitude between two freely floating test-masses to an accuracy of picometer and nanorad, respectively, which makes it the most precise metrology system for space as of today\cite{Hei2003},\cite{Aud2011}.
Its measurement data allow determining the residual differential acceleration between the two test-masses to an accuracy of better than $3\times 10^{-14} {\rm m/s^2}/\sqrt{Hz}$ (2 orders of magnitude better than state-of-the-art accelerometers). The OMS output also feeds into the {\it drag-free attitude and control system} (DFACS) which counteracts external perturbations acting on the spacecraft and test-masses and therefore allows demonstration of quasi-free floating test-masses.

As an overview we give a detailed account of the working principles of the Optical Metrology System which is essentially based on the digital phase-readout and processing of the signals of four heterodyne interferometers located on an ultra-stable optical bench.
The impact of various noise sources on the overall measurement performance is briefly discussed, among which frequency and power fluctuations of the laser source, parasitic optical sidebands induced by RF-crosstalk in the modulator and phase noise combined with path-length variations induced by signal transmission through optical fibres are the most prominent ones. These limitations can be overcome by the use of digital control loops which suppress laser frequency noise and differences in optical path-length to such a degree that the final performance is fundamentally only limited by the sensor and quantization noise of the digital phase-meter. Furthermore, overall stability and performance is much improved by balancing the interferometer arm-lengths to a good degree, which specifically reduces the effect of laser frequency fluctuations.

An important feature of the general system design is that the phase of the {\it reference interferometer} is subtracted from the phase of the test-mass {\it position interferometer}, thereby greatly suppressing common mode noise fluctuations occurring in volatile optical path sections shared by both interferometers.
Analyses of experimental data recorded during recent test and qualification campaigns point to an as yet unaccounted phenomenon which relates to imperfect common mode noise suppression when the two interferometer phases are subtracted in the digital domain. We find that the efficiency of common mode noise subtraction depends strongly on the relative difference of the interferometer phases: Increasing phase differences lead to decreasing common mode noise rejection. We theoretically investigate the expected level of noise rejection and discuss its dependency on the relative phase and on the amplitude of the phase-noise whilst comparing our findings to those observed in experimental measurements. These investigations are directly applicable to similar metrology systems where a highly stable phase is obtained by suppressing the common mode noise through subtraction of a reference phase.

\section{Overview over the Optical Metrology System}
\subsection{\label{section:overview} Working principle and System Design}
\begin{figure}
   \begin{center}
   \begin{tabular}{c}
   \includegraphics[height=9cm]{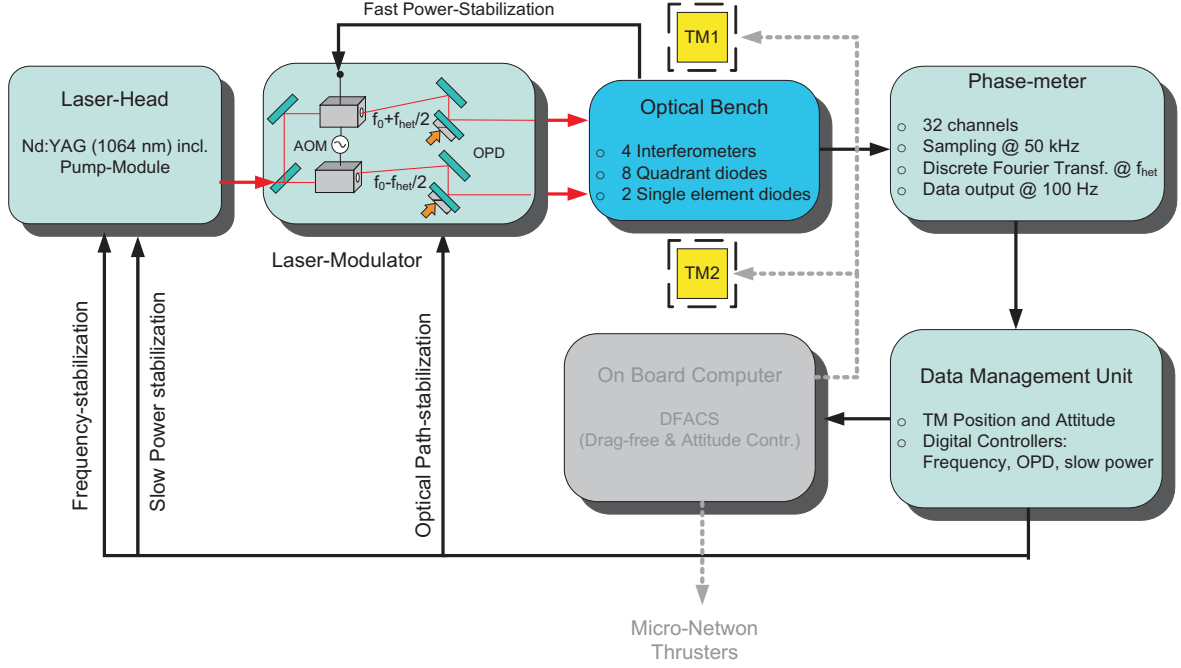}
   \end{tabular}
   \end{center}
   \caption[example]
   { \label{fig:oms_schematic}
Schematic of the working principle of the optical metrology system of LISA Pathfinder.}
\end{figure}

The Optical Metrology System (OMS) is one of the core subsystems of the LISA Technology Package (LTP), the scientific payload of LISA Pathfinder (LPF). LPF is the technological precursor mission to LISA, the gravitational wave detector in space. Whereas in LISA laser beams propagate over a distance of 5 million km between three spacecraft forming a triangular constellation, in LPF the distance between freely floating test-masses is shrunk to 38 cm where the arm-length is far too small to actually detect gravitational waves. Considering that the two test-masses are replaced by dummy mirrors for ground testing many aspects of the setup resemble those of conventional interferometers for high precision metrology.
The OMS essentially comprises the following units:
\begin{itemize}
  \item Laser unit with heterodyne modulator
  \item Optical bench with 4 heterodyne interferometers
  \item Phase-meter and quadrant diodes for read-out of the interferometer phases
  \item Data-Management Unit (DMU) for processing the phase data and determining test-mass position and attitude
\end{itemize}
A short schematic over the working principle of the optical metrology system is given in Fig.\ref{fig:oms_schematic}.
The output of the Nd:YAG laser unit (at 1064 nm) is split at the input of the modulator into {\it measurement} and {\it reference beam}, where the name of the former derives from the fact that part of it reflects off the test-mass and therefore contains information on its position, which differ by the heterodyne frequency $f_{\rm het}=$ 1 kHz. The modulator additionally contains piezo-actuated prisms for stabilizing the relative Optical Path-length Difference (OPD) between the two beams which are then guided onto the optical bench (schematically drawn in Fig.\ref{fig:opt_bench}) through two single-mode fibres. The optical bench comprises 4 interferometers, two of which -the x1 and x12 interferometer- measure the position of test-mass 1 and 2. The output phase of the reference interferometer $\phi_R$ is subtracted from the output phases $\phi_1$ and $\phi{12}$ of the position interferometers (this process shall be referred to as {\it phase-adjustment}) to suppress common mode phase noise. Whereas the other three interferometers have balanced (equal length) arms, as counted from the splitting point S between measurement and reference beam onwards, the arm-lengths of the frequency interferometer are deliberately mismatched by 40 cm to convert frequency fluctuations $\delta\nu$  into phase fluctuations $\delta\phi$ which scale proportional to the arm-length difference L: $\delta\phi=2\pi{\rm L}/c~\delta\nu$.

The interferometer phases are detected by quadrant diodes and the channel signals received and processed by the phase-meter. The latter samples the signals $S(t)$ at $f_{\rm samp}=50$ kHz and performs a single-bin Discrete Fourier Transform (DFT) at the heterodyne frequency $f_{het}= 1$ kHz for a period T=10 ms. The data, consisting of the two complex components of the phase vector $F_t$, are transmitted to the Data Management Unit (DMU) for further processing at the update rate $f_{up}=100$ Hz:
\begin{eqnarray}
\Re(F_t)=\sum_{k=0}^{N-1} S(k\cdot\Delta t) \cos\left(\frac{2\pi m k }{N}\right)\nonumber\\
\Im(F_t)=\sum_{k=0}^{N-1} S(k\cdot\Delta t) \sin\left(\frac{2\pi m k }{N}\right),\label{eqn:phase_int}
\end{eqnarray}
where $N=$500 is the number of points in the DFT and the heterodyne frequency corresponds to bin number 'm': $f_{het}=m~2\pi/T$.
In the DMU the test-mass position and attitude are calculated, the former through averaging the phases of four quadrants and tracking the resulting phase and the latter through finding the phase difference between two halves of a diode. Due to bandwidth constraints of the MilBus data bus connecting the various electronic units, the data are down-sampled to 10 Hz by application of a moving average filter before being transmitted to the Drag-Free Attitude and Control System (DFACS). The DFACS controls the spacecraft using micro-Newton thrusters as actuators and uses electrostatic forces/torques to control test-mass degrees-of-freedom in a way that the freely-floating test-masses remain centered between the electrodes\cite{Fic2008}. The DMU also implements digital controllers for laser frequency control, Optical Path-length Difference (OPD) stabilization, and slow power stabilization \cite{Hec2011}. The latter was found to be necessary although the interferometer is designed to be robust against phase fluctuations of volatile components (see Fig.\ref{fig:opt_bench} in next section) due to the occurrence of optical sidebands which arise from the cross-coupling of RF-signals going to the AOMs and lead to significant performance degradation if the OPD is left unstabilized\cite{Wan2006}. The frequency stabilization scheme uses a fast and a slow loop in cascaded configuration, where piezo actuators on the NPRO resonator and a temperature controller are used as actuators,
respectively. A fast analogue power stabilization loop (bandwidth of tens of kHz) uses the two accousto-optic modulators (AOMs) of the heterodyne modulator as actuators.

\subsection{Optical Bench and Interferometers}
\begin{figure}
   \begin{center}
   \begin{tabular}{c}
   \includegraphics[width=16cm]{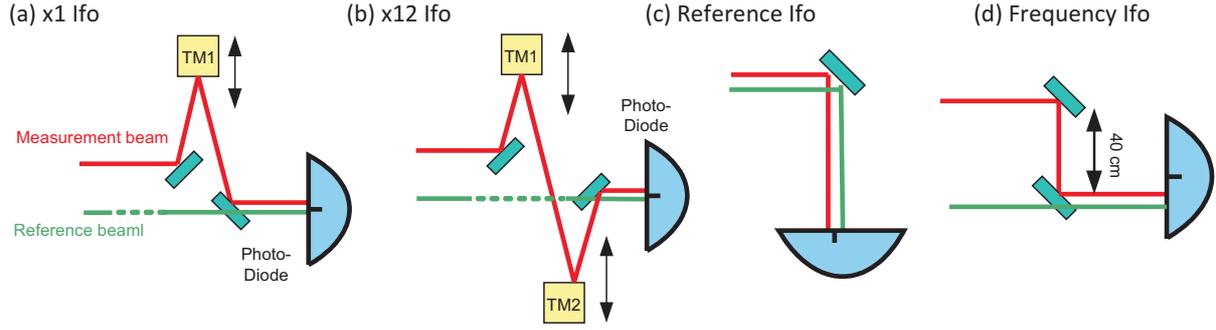}
   \end{tabular}
   \end{center}
   \caption[example]
   { \label{fig:interferometer_comparison}
Basic schematic (abstract) of the 4 interferometer types used on the optical bench.}
\end{figure}

\begin{figure}
   \begin{center}
   \begin{tabular}{c}
   \includegraphics[height=10cm]{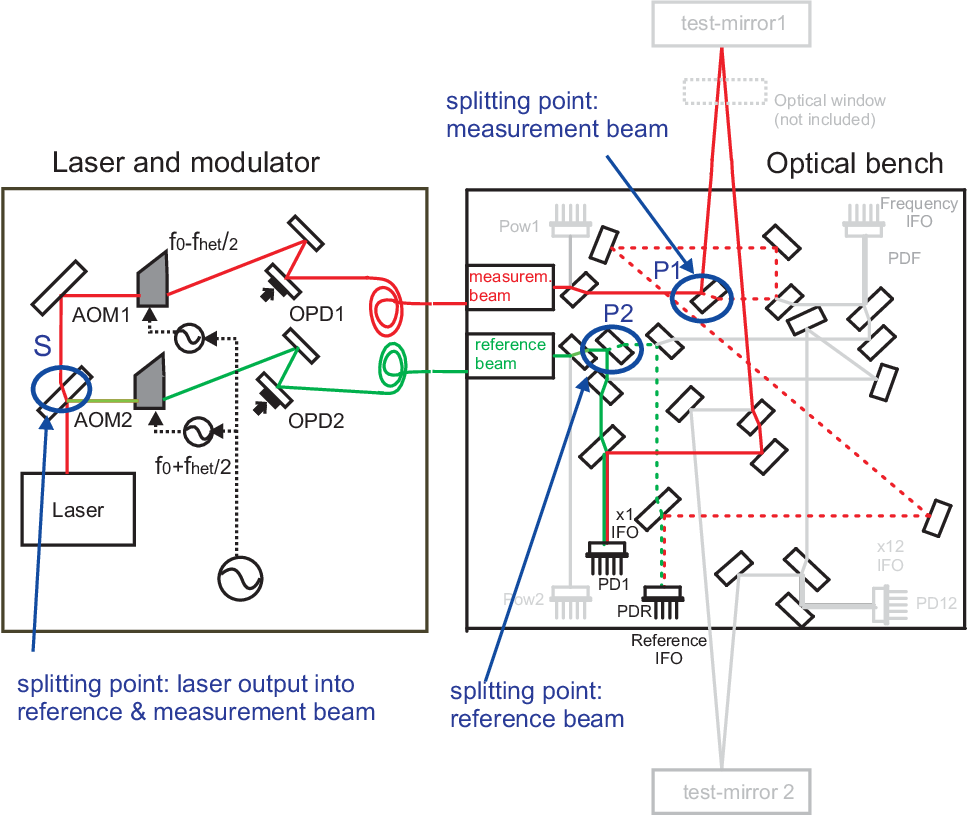}
   \end{tabular}
   \end{center}
   \caption[example]
    { \label{fig:opt_bench}
The optical bench of the OMS is drawn to scale. Beams of the "x1" and "reference" interferometers are shown as solid red/green and broken red/green lines, respectively, other beams are greyed out. Interferograms are detected by quadrant diodes PD1 and PDR, respectively.}
\end{figure}

 The optical bench accommodates 4 interferometers\cite{Hei2003} for which the abstracted schematics and basic design features of are shown in Fig.\ref{fig:interferometer_comparison} and may be summarized as follows:
\begin{itemize}
  \item {\bf Science Interferometer x1:} It measures the longitudinal displacement along one axis and the orientation around two axes of test-mass 1.
  \item {\bf Science Interferometer x12:} It measures the relative longitudinal displacement along one axis and the relative orientation around two axes between test-mass 1 and test-mass 2.
  \item {\bf Reference Interferometer R:} The output of the reference interferometer (reference phase $\phi_R$) is subtracted from the output of all other interferometers so that common mode noise is suppressed. Additionally, $\phi_R$ is used to stabilize the optical path-length difference (OPD) between measurement and reference beam to mitigate the effect of unwanted optical sidebands \cite{Wan2006}.
  \item {\bf Frequency Interferometer F:} The optical path-length between measurement and reference beam is deliberately mismatched by L=40 cm so that laser frequency fluctuations $\delta\nu$ are converted into phase fluctuations $\delta\phi$ according to: $\delta\phi=2\pi L/c~\delta\nu$. Therefore, the output phase exhibits increased phase fluctuations which serve as feedback to the frequency control loop.
\end{itemize}
A basic schematic of the laser and its modulator as well as the optical bench comprising the 4 interferometers is given in Fig.\ref{fig:opt_bench}.
Here only the beam paths for the measurement interferometer x1 (solid red and green lines) and for the reference interferometer R (broken red and green lines) are drawn in color, the beams of the other interferometers are not discussed any further and therefore greyed out. The measured phase of every interferometer is based upon the interference between measurement and reference beams which are split from another at the laser output beam splitter {\it S}. Both beams are further divided into the arms of the various interferometers at beam splitters {\it P1} and {\it P2}, respectively.

However, the volatile optical path section from {\it S} to {\it P1} for the {\it measurement beam} and from {\it S} to {\it P2} for the {\it reference beam} is shared between all interferometers so that it is equally reflected in the respective output phases. Therefore, common mode phase noise occurring in these path sections is expected to be removed entirely when the phase of the reference interferometer is subtracted from the phase of any other interferometer. Phase fluctuations occurring in the other path sections which are specific to each interferometer may be neglected as these path sections are located on the ultra-stable optical bench. The latter only comprises optical elements made from fused silica which are hydroxyl-catalysis bonded\cite{Ell2005} to the Zerodur baseplate so that a quasi-monolithic structure with superior stability and negligible sensitivity to thermal expansion is formed.

\subsection{Longitudinal and angular measurements}
\begin{figure}
   \begin{center}
   \begin{tabular}{c}
   \includegraphics[width=16cm]{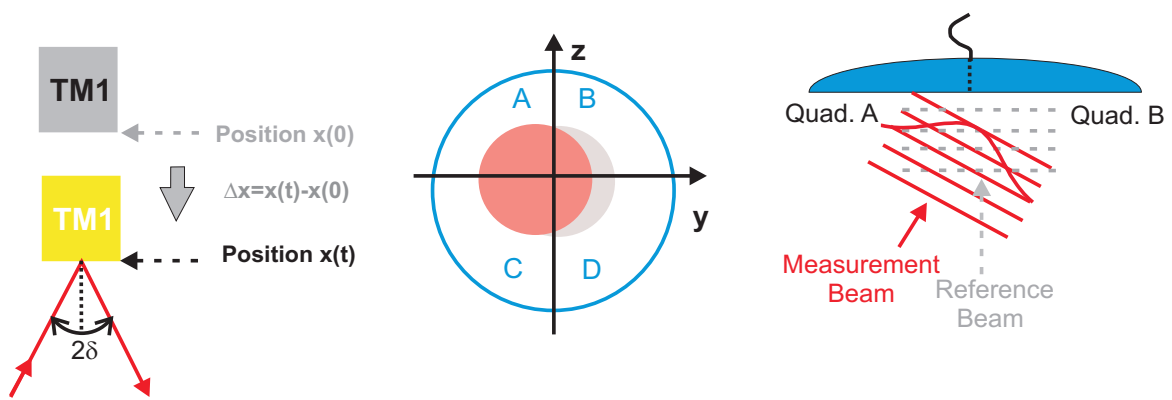}
   \end{tabular}
   \end{center}
   \caption[example]
    { \label{fig:measurement_types} The 3 types of measurement performed with the science interferometers x1 and x12: (a) The longitudinal displacement is determined from the Ifo-phase. (b) The test-mass tilt is determined from the power difference between diode quadrants due to the displacement of the incident beam (DC-measurement). (c) The test-mass tilt is determined from the phase difference between diode quadrants due to the inclined wavefront of the incident beam (DWS-measurement).}
\end{figure}

The optical metrology system performs 3 types of measurement which are schematically displayed in Fig.\ref{fig:measurement_types}.
On the one hand, the longitudinal test-mass displacement $\Delta x$ can be directly determined from the interferometer phase change $\Delta\phi$ through the relation
\begin{equation}
\Delta x=\frac{\lambda}{4\pi\cos\delta}\Delta\phi,
\label{eq:position_conversion}
\end{equation}
where $\lambda$=1064 nm is the laser wavelength and $\delta$=4.5 deg is the acute incidence angle of the beam on the test-mass (a). On the other hand, angular tilts of the test-masses can be determined from either measurements of the power difference (b) or measurements of the phase difference (c) between diode quadrants.

The power difference signal ${\rm DC}_{\phi}$ relates to the test-mass tilt angle $\phi$ though the following equation:
\begin{eqnarray}
{\rm DC}_{\phi}&=&\frac{P_{\rm left}-P_{\rm right}}{P_{\rm left}+P_{\rm right}}\approx K_{\rm DC}\times \phi\nonumber\\
K_{\rm DC}& \approx &\frac{P_m}{P_m+P_r}\sqrt{\frac{2}{\pi}}\frac{4 L_{\rm tm}}{w_{\rm m}},\nonumber\\
\label{equ:K_DC}
\end{eqnarray}
where $L_{\rm tm}$ denotes the distance between test-mass and photo-diode and $w_m$ the beam waist of the measurement beam.
The phase difference signal ${\rm DWS}_{\phi}$ is found from the complex phase-vectors ($F_i$, i=A,B,C,D; see Fig.\ref{fig:measurement_types}) of the diode quadrants which are determined from the discrete Fourier transform at the heterodyne frequency. Through this technique, which is also referred to as Differential Wavefront Sensing (DWS)\cite{Hei2003}, the test-mass tilt angle $\phi$ is determined from the DWS-signal as follows\cite{Hec2010b}:
\begin{eqnarray}
{\rm DWS}_{\phi}&=& {\rm arg}\left\{\frac{F_A+F_C}{F_B+F_D}\right\}\approx K_{\rm DWS}\times\phi\nonumber\\
K_{\rm DC}&\approx& \sqrt{2\pi}\frac{w_{\rm eff}}{\lambda}\left(1-\frac{L_{\rm tm}}{R}\right),\nonumber\\
\label{equ:K_DWS}
\end{eqnarray}
where $w_{\rm eff}$ is the average beam waist of measurement and reference beam and $R$ the radius of curvature of the measurement beam.
Inserting the typical design parameters of the optical metrology system into Eqs.\ref{equ:K_DC},\ref{equ:K_DWS}, one finds that the theoretical predictions are in excellent agreement with calibration measurements which yielded coupling constants of $K_{\rm DC}\approx 300$ and $K_{\rm DWS}\approx 5000$.

\section{Interferometer Performance}
\subsection{Phase-meter Performance and Quantization Noise }
When analogue signals are digitized, they are affected by the so-called {\it quantization error}. If during the digitization process the internal signal representation is rounded off to the nearest integer, the probability distribution $p_{qn}$ can be modeled as a random variable that is uniformly distributed between -LSB/2 and +LSB/2, where LSB stands for the least significant bit. One then finds for the RMS-value of the quantization noise fluctuations $n_{\varphi}$\cite{Gra2003}:
\begin{equation}
\langle n_{\varphi}^2\rangle =\sqrt{\langle x^2 p_{qn}\rangle}={\rm LSB}\frac{1}{\sqrt{12}}=\frac{{\rm FD}}{2^n}\frac{1}{\sqrt{12}},
\label{equ:LSB_QN}
\end{equation}
where {\it n} is the number of bits and {\it FD} is the full dynamic range of the ADC.
The signal-to-quantization-noise ratio (SQNR) is the ratio between the RMS amplitude of the fundamental input frequency to the RMS amplitude of the quantization noise so that we find for a sinusoidal input signal which covers the full dynamic range (amplitude of FD/2) of the ADC:
\begin{equation}
{\rm SQNR_{dB}}=20~\log_{10}\left(\frac{{\rm FD}/\sqrt{8}}{{\rm FD}/\sqrt{12}}\cdot \frac{1}{2^n}\right)=(6.02\cdot n+1.76) dB
\end{equation}
However, in an actual ADC harmonic distortions and other imperfections come into play in addition to the quantization noise so that it is useful to introduce another quantity, the signal-to-noise plus distortion ratio (SINAD). The SINAD is defined as the ratio between the RMS amplitude of the fundamental input frequency to the RMS amplitude of all other frequency components at the A/D output, where the output is band-limited to frequencies from DC to below half the sampling frequency.
In that context, the effective number of bits (ENOBs) $n_{\rm eff}$ is a direct measure of ADC resolution and relates to the SINAD through the relation
\begin{equation}
n_{\rm eff}=\left({\rm SINAD_{dB}}-1.76\right)/6.02,
\end{equation}
For the 16-bit ADC used in the OMS (LTC1604 by Linear Technology) the manufacturer specifies a SINAD of 90 dB and therefore an effective number of 14.7 bits for our sampling frequency of  50 kHz.
As described by Eq.\ref{eqn:phase_int}, the output-phase is found by multiplication of the input signal with a reference oscillation and integration over a multiple of the oscillation period. Consequently, the quantization error distorts the measured output phase . For a fixed phase relationship between input and reference signal this leads to output phase fluctuations given by Eq.\ref{equ:quant_noise_main} which is derived in appendix A:
\begin{equation}
\Delta\phi_{\rm qn}=\frac{1}{2^{n-1}\sqrt{3N_F}},
\label{equ:quant_noise_main}
\end{equation}
where $N_F=f_{\rm samp}/f_{\rm het}=50$ is the number of sampling points in one heterodyne period.
For the typical parameters used in the LISA Pathfinder phasemeter ($f_{\rm samp}$=50 kHz, $f_{\rm het}$=1 kHz, $f_{\rm up}$=100 Hz) and using an effective number of $n_{\rm eff}=14.7$ bits, this yields for the linear spectral density $LSD(\phi_{\rm qn})\approx 1\mu{\rm rad}/\sqrt{Hz}$. This value is somewhat lower than what was measured for the individual phase-meter channels, which indicates that the phase-meter performance is limited by the noise from the internal electronics rather than ADC noise.

\subsection{Noise Sources of the Interferometer}
\begin{figure}
   \begin{center}
   \begin{tabular}{c}
   \includegraphics[width=12cm]{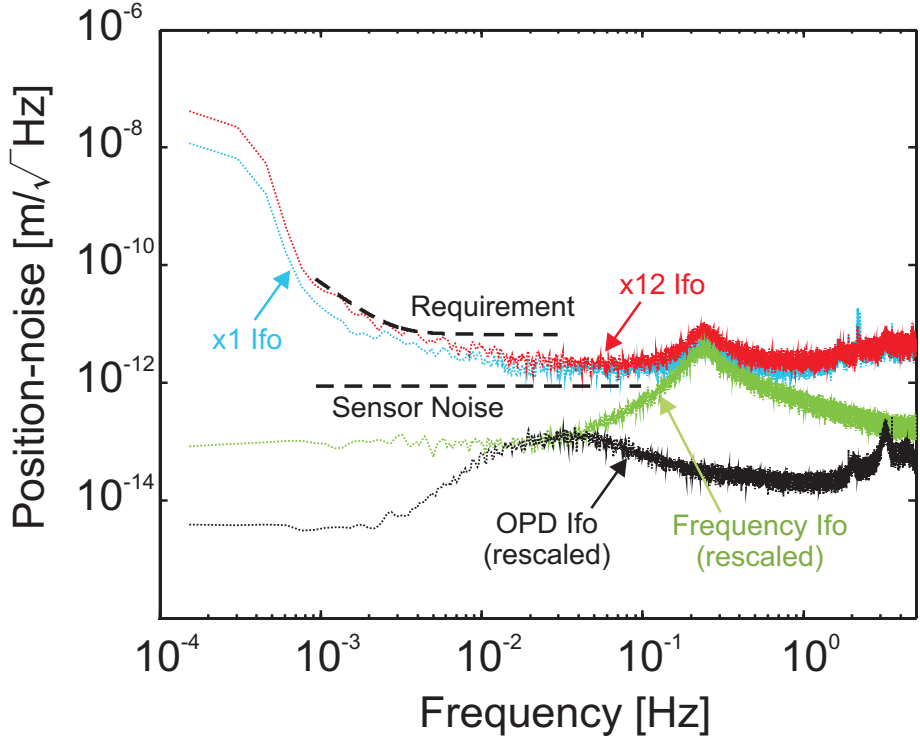}
   \end{tabular}
   \end{center}
   \caption[example]
   { \label{fig:x1_performance}
The performance curves for the x1 and x12 iterferometer are given by the blue and red lines, respectively. The dashed lines denote the measurement requirement and the achievable optimum (sensor limit). The noise contributions due to frequency fluctuations and OPD-variations are given by the green and black lines, respectively.  }
\end{figure}

The major focus of this paper lies on the measurement performance and the impact of the various noise sources on the latter. In order to comply with the overall measurement accuracy requirement for the LISA Pathfinder mission, the linear spectral density (LSD) of the OMS position measurement noise is required to be below 6.4 pm in the measurement band between 1 mHz and 30 mHz according to the following equation:
\begin{eqnarray}
&&{\rm LSD}(x)= 6.4\times 10^{-12}\sqrt{1+\left(\frac{f}{3~{\rm mHz}}\right)^{-4}}\frac{m}{\sqrt{Hz}}\nonumber\\
&& 1~{\rm mHz}\leq f \leq 30~{\rm mHz}\nonumber\\
\end{eqnarray}
In order to obtain a measurement performance plot, the phase data from the science interferometers (after appropriate conversion into position according to Eq.\ref{eq:position_conversion}) are recorded for a period of several hours and then Fourier transformed, after removing the overall trend and using an appropriate window function, to obtain the Linear Spectral Density (LSD) of the position noise. The OMS measurement performance is affected by the following noise sources
\begin{enumerate}
  \item Laser frequency fluctuations
  \item Laser power fluctuations
  \item Sensor noise (ADC, electronics, quantization)
  \item Phase noise (vibrations, temperature fluctuations, optical fibres, AOMs ...)
\end{enumerate}
As already discussed in section \ref{section:overview}, digital control loops were implemented for laser frequency stabilization and stabilization of the optical path-length difference (OPD)\cite{Hec2011}. Consequently, the phase fluctuations arising from frequency noise coupled with imperfectly balanced arm-lengths of the science interferometer could be suppressed to a negligible level. This fact is indicated by the green curve in Fig. \ref{fig:x1_performance} which was obtained from recorded data of the frequency interferometer where the phase measurements were scaled by the approximate ratio of arm-length differences between science and frequency interferometer. It is shown to lie well below the measurement performance curves of the x1 and x12 science interferometers which are denoted by the blue and red curves, respectively.
Similarly, the performance degrading effects of optical side-bands (from modulator cross-talk) were mitigated through the OPD stabilization scheme whose contribution to the science interferometer performance is shown by the black curve of Fig. \ref{fig:x1_performance}. The latter curve was obtained from a re-scaled phase noise measurement of the OPD interferometer where previously measured amplitude ratios of the optical sidebands \cite{Wan2006} were considered in the transformation. OPD-related phase noise was therefore also shown to be negligible. Note that the strong increase of the noise floor for the x1 and x12 performance curves at frequencies below 1 mHz is due to temperature fluctuations which cause the test-mirrors to move and the interferometer phases to change accordingly. Atmospheric disturbances were generally avoided by performing the measurements under vacuum conditions.

Laser power was stabilized to a level on the order of $10^{-6}/{\sqrt Hz} <\delta P/P<10^{-5}/ {\sqrt Hz}$ through a fast analogue power control loop which acts on the AOMs of the laser modulator and operates at a bandwidth of 50 kHz. A digitally implemented slow power loop (acting on the laser diode current) defines the output power setpoint and is tasked with offloading the actuators of the fast loop to avoid saturation. Power fluctuations are converted into phase-noise $n_{\phi}$ through the action of the phase-meter according to the simple expression\cite{Hec2013}: $n_{\phi}=\delta P/(P\sqrt{2})$. Therefore, when the power stabilization scheme is active, the largest power fluctuations give rise to phase fluctuations on the order of only a few micro-rad. These correspond to negligible position fluctuations well below 1 pico-meter.

Although the control loops aim at suppressing adverse noise sources as much as possible, a fundamental limit persists that cannot be overcome, namely the sensor-noise relating to signal sampling and processing by the phase-meter. The sensor noise was found to be dominated by noise from the phase-meter electronics which exceeds quantization and ADC noise\cite{Hoy2003}. It was further found to scale inversely proportional with the amplitude of the heterodyne signal up to a maximum of half the full dynamic range where sensor noise was as low as $2~\mu {\rm rad}/\sqrt{Hz}$. For the performance measurements of Fig.\ref{fig:x1_performance} the signal dynamics covered only 12 percent of the full range and the sensor noise reached a value of $10~\mu {\rm rad}/\sqrt{Hz}$, corresponding to position noise of less than $1~{\rm pm}/\sqrt{Hz}$. However, we observed that the performance of position measurements, given by the blue and red curves in Fig.\ref{fig:x1_performance}), was generally worse than the sensor limit so that the overall performance is limited by a different noise source rather than the sensor noise. Furthermore, the performance varied between measurements, sometimes the x1 performance being better than the x12 performance, and sometimes the other way round. Questions on the possible origin of the limiting noise therefore arise.

 \subsection{High Frequency Noise Rejection}
At this point it is important to note that  other noise sources have only been suppressed within the bandwidth of the servo loops (typically operating at 100 Hz) so that high frequency phase noise at or above $f_{\rm het}$ still couples into the phase-meter input signals without attenuation. Although the interferometers are designed to reject common mode noise at low frequencies through subtraction of a reference phase, this does not hold for high frequency components of the noise spectrum, in particular those components around the heterodyne frequency.
In a recent paper we discussed how high frequency amplitude and phase fluctuations of the input signals degrade the measurement performance\cite{Hec2013}. We showed that the efficiency of common mode noise subtraction in the digital domain depends on the relative phase between the two input signals and found that increasing phase differences lead to decreasing common mode noise rejection. In the performance measurements depicted in Fig.\ref{fig:x1_performance} one input signal with phase $\phi_1$ is derived from the science interferometer x1 from which we subtract the signal from the reference interferometer with phase $\phi_R$.
\\The characteristic features of each noise type as well as the respective common mode rejection properties are summarized in Tab.\ref{tab:noise_types}, which was constructed based on the derivations given in Appendix A of this paper and those given in \cite{Hec2013}.
From Tab.\ref{tab:noise_types} we see that if high frequency amplitude noise is dominant, the output noise of the signal difference scales as $\sin[(\phi_1-\phi_R)/2]$, reaching a maximum for $\phi_1-\phi_R=180$ deg. If, on the hand, high frequency phase noise dominates, the output noise of the signal difference is proportional to $\sin[\phi_1-\phi_R]$, reaching a maximum for $\phi_1-\phi_R=90$ deg.
\begin{table}[h]
\caption{The linear spectral density (LSD) of the output phase noise is given for a single signal of phase $\phi_1$ (first row) or a signal difference of phase $\phi_1$-$\phi_R$ (second row) which are affected by various noise sources. The parameter $n_{ld}$ denotes the linear spectral density of the input signal which is expressed in ${\rm rad}/\sqrt{Hz}$ for phase noise and ${\rm Volt}/\sqrt{Hz}$ for amplitude noise. The parameter $A_{\rm in}$ denotes the amplitude of the input signal in Volt.}
\label{tab:noise_types}
\begin{center}
\begin{tabular}{l|lll} 
\rule[-1ex]{0pt}{3.5ex}  &quantization noise & amplitude noise & phase noise  \\
\hline
\rule[-1ex]{0pt}{3.5ex}  $\vphantom{\frac{\frac{N}{N}}{\frac{N}{N}}}$LSD($\phi_1$)         & $\frac{\sqrt{2}}{2^n\sqrt{3N_F f_{up}}}$ & $\frac{n_{ld}\sqrt{2}}{A_{in}}$ & $\approx\frac{n_{ld}}{\sqrt{2}}$   \\
\rule[-1ex]{0pt}{3.5ex}  $\vphantom{\frac{\frac{N}{N}}{\frac{N}{N}}} $LSD($\phi_1$-$\phi_R$)& $\frac{2}{2^n\sqrt{3N_F f_{up}}}$ & $\frac{n_{ld}2\sqrt{2}}{A_{in}}\left|\sin\left(\frac{\phi_1-\phi_R}{2}\right)\right|$& $n_{ld}\sqrt{2}\left|\sin\left(\phi_1-\phi_R\right)\right|$  \\
\end{tabular}
\end{center}
\end{table}

For the performance plots of Fig.\ref{fig:x1_performance} the average phase differences are $\phi_1-\phi_R=27$ deg for the x1 interferometer and $\phi_1-\phi_R=33$ deg for the x12 interferometer, which explains why the performance of the x1 interferometer is slightly better than the one for the x12 interferometer.
\begin{figure}
   \begin{center}
   \begin{tabular}{c}
   \includegraphics[width=12cm]{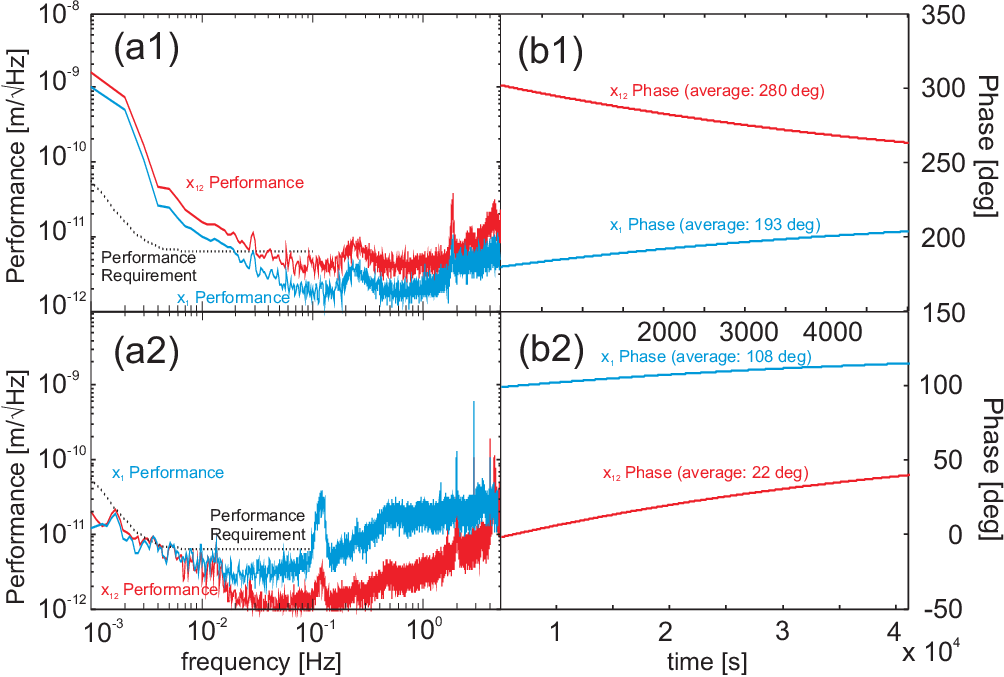}
   \end{tabular}
   \end{center}
   \caption[example]
   { \label{fig:case1}
 (a1) Performance measurement for the x1 (blue) and the x12 (red) interferometer of the optical bench (engineering model). (b1) The interferometer phase is plotted against time for the same measurement period which was evaluated in (a1). The performance for x1 is better than for x12 because its average phase is close to 180 degrees. (a2) Performance measurement for the x1 (blue) and the x12 (red) interferometer of the optical bench (flight model). (b2) The interferometer phase is plotted against time for the same measurement period which was evaluated in (a2). The performance for x12 is better than for x1 because its average phase is close to 0 degrees. }
\end{figure}
We analyzed a variety of different data sets and found that imperfect common mode noise rejection of high frequency phase noise could very well be ascribed to the qualitative explanation of the observed features. Figure \ref{fig:case1}(a1) plots the science interferometer performance curves for a certain measurement and (b1) shows the corresponding curves of the average phases $\phi_1$ and $\phi_{12}$. The average phases drift over a range of approximately 30 deg during the measurement due to temperature induced test mirror movements whilst the reference phase $\phi_R$ remains stabilized to 0 deg. As expected for phase noise, the performance for x1  is better because its average phase is only 13 deg away from the minimum (at 180 deg) as compared to the performance of x12 where the average phase is 100 deg away from the minimum and therefore close to worst case (at 270 deg).
The opposite scenario is depicted in Figs.\ref{fig:case1}(a2) and (b2). Here, the average phase of x12 is only 22 deg away from zero and therefore yields a better performance than the one for x1, where the phase is 108 deg  away from the minimum.

\section{Conclusions}
In this paper we discussed that several conventionally referenced noise sources (laser frequency noise, laser power fluctuations, optical path-length difference noise, and sensor noise) cannot explain the typically achieved measurement performance and the variability of certain characteristic features thereof for the LISA Pathfinder optical metrology system.
The interferometer design is such that common mode noise is generally well rejected by the subtraction of a reference phase from the actual measurement observable, but this only works at low frequencies. High frequency phase and amplitude noise were shown to couple into the low frequency measurement band [1 mHz, 30 mHz] as common mode noise rejection deteriorates with increasing phase difference between measurement and reference signal. The data from several measurement campaigns seem to point to high frequency phase noise as a possible origin of the measured performance limits.

\acknowledgments     
The authors gratefully acknowledge support from the {\it Deutsches Zentrum f\"ur Luft- und Raumfahrt} (DLR) and the European Space Agency (ESA) for performing the test campaigns with the optical metrology system (Engineering Model and Flight Model) from which some data are referenced in this paper. The data for Figures \ref{fig:case1}(a1) and (b1) were recorded during tests on the premises of the Albert Einstein Institute in Hannover with the support of AEI scientists, the data for (a2) and (b2) were recorded at IABG in Munich during the LISA Pathfinder thermal-vacuum test campaign which were organized and supported by Astrium UK and the European Space Agency (ESA), respectively. We would like to thank David Hoyland (University of Birmingham), Gerhard Heinzel (Albert Einstein Institute), Vinzenz Wand (now OHB Systems), Paul McNamara and Bengt Johlander (European Space Agency), Domenico Gerardi, Nico Brandt, and Ulrich Johann (Astrium Germany) for fruitful discussions. Special thanks go to R\"udiger Gerndt (LTP System engineer) for expert advice, many consultations and help to improve the manuscript.

\section*{APPENDIX A: Output phase fluctuations from quantization noise}
In the following we shall derive an expression for the output phase noise that is caused by the signal quantization.
A sinusoidal input signal $S(k\Delta t)$ is quantized by an ADC with 'n' bits as described below:
\begin{equation}
S(k\Delta t)_q=2^{-(n-1)}{\rm round} \left( 2^{n-1}S(k\Delta t)\right)
\label{equ:quant_relation}
\end{equation}
The analogue input signal $S$ relates to the quantized output signal $S_q$ through the relation
\begin{equation}
S(k\Delta t)=S_q(k\Delta t)+n_k,\label{equ:qnt_rounding}
\end{equation}
where $n_k$ denotes the quantization noise arising from the rounding errors which depends on the sampling step $k$.
Using Eq.\ref{equ:qnt_rounding} and considering that the input signal is sinusoidal with a phase offset '$\phi$'we rewrite Eq. \ref{eqn:phase_int} to read:
\begin{eqnarray}
\Re(F_t)=\sum_{k=0}^{N-1} S(k\cdot\Delta t) \cos\left(\frac{2\pi m k }{N}\right)\nonumber=\sum_{k=0}^{N-1}\left[A_{\rm in} \cos\left(\frac{2\pi m k }{N}-\phi\right)+n_k\right]\cos\left(\frac{2\pi m k }{N}\right)\\
\end{eqnarray}
where '$A_{in}$' is the amplitude of the input signal. We neglected the quantization noise for the second cosine term which is assumed to be stored in the Phase-meter lookup table at a sufficiently high precision (say $>20$ bits) such that it can be omitted from further considerations.
Expanding the first cosine term in the above expression we find:
\begin{eqnarray*}
\Re(F_t)&=&\sum_{k=0}^{N-1}A_{\rm in} \left[\cos\left(\frac{2\pi m k }{N}\right)\cos\phi+\sin\phi\sin\left(\frac{2\pi m k }{N}\right)+n_k\right]\cos\left(\frac{2\pi m k }{N}\right)\\
&=&\frac{A_{\rm in} N}{2}\cos\phi+\sum_{k=0}^{N-1}n_k\cos\left(\frac{2\pi m k }{N}\right)
\end{eqnarray*}
A similar expression can be found for $\Im(F_t)$ so that from the ratio of the two expressions the phase $\phi$ is found:
\begin{equation}
\tan\phi_{qn}=\frac{\Im(F_t)}{\Re(F_t)}=\frac{\sin\phi+\frac{2}{NA_{\rm in}}\sum_{k=0}^{N-1}n_k\sin\left(\frac{2\pi m k }{N}\right)}{\cos\phi+\frac{2}{NA_{\rm in}}\sum_{k=0}^{N-1}n_k\cos\left(\frac{2\pi m k }{N}\right)}\label{equ:full_noise}
\end{equation}
We note that it is helpful to visualize expression \ref{equ:full_noise}  in the complex plane as follows: The tip of a phase vector lies on the unit circle with angle $\phi$ towards the real axis. Around the tip of this phase a number $N$ of small noise vector contributions perform a random walk in 2 dimensions. Without loss of generality the coordinate system can always be transformed in a way that the phase vector is parallel to the real axis ($\phi=0$). Then the vector sum of the noise components perpendicular to the real-axis (proportional to $\sin(2\pi m k/N)$ determines the fluctuations of the phase $\phi_{\rm qn}$ of the total phase vector whereas the parallel components (proportional to the cosine) can be neglected (second order effect).
We find for the mean-square fluctuations:
\begin{equation}
\langle \delta\phi_{\rm qn}^2 \rangle=\frac{4}{N^2A_{\rm in}^2}\sum_{k=0}^{N-1} \langle n_k^2\rangle\sin^2\left(\frac{2\pi m k }{N}\right)
\label{equ:fluctuations}
\end{equation}
After inserting Eq. \ref{equ:LSB_QN} into Eq. \ref{equ:fluctuations} and executing the sum we obtain for the measured phase fluctuations induced by quantization noise
\begin{equation}
\Delta\phi_{\rm qn}=\sqrt{\langle\delta\phi_{\rm qn}^2\rangle}=\frac{\sqrt{2}}{2^{n}\sqrt{3N}},
\label{equ:quant_phase_noise}
\end{equation}
which has already been derived in \cite{Hoy2003} from a similar ansatz.
At this point we must stress that quantization noise is somewhat more complex as it only manifests its statistical properties if the phase between input and reference signal does not remain fixed. Otherwise the random fluctuations are exactly identical from period to period and therefore all periods correlate with another.
Assuming such a rigid lock between the phase of input and reference signal we obtain after some algebra for the measured phase fluctuations
\begin{equation}
\Delta\phi_{\rm qn}({\rm phase-locked})=\frac{1}{2^{n-1}\sqrt{3N_F}}=\sqrt{\frac{2N}{N_F}}\Delta\phi_{\rm qn},
\label{equ:quant_noise_fixed}
\end{equation}
where $N_F=f_{\rm samp}/f_{\rm het}=50$ is the number of sampling points in one heterodyne period. We observe that the noise is increased by the square root of the number of heterodyne oscillations within one FFT period ($N/N_F$) as well as by a factor $\sqrt{2}$ (which comes from the correlation between the two halves of one oscillation period) with respect to Eq.\ref{equ:quant_phase_noise}.
The predictions of Eq.\ref{equ:quant_noise_fixed} were compared to the results of numerical simulations where the input signal was quantized and the phase measured according to Eq.\ref{eqn:phase_int} and excellent agreement was found. In the simulations the input signal phase was incrementally changed from 0 to pi and the output phase fluctuations were averaged over all phase steps.

\end{document}